\documentclass[aps,pre,preprint,groupedaddress,showpacs]{revtex4-1}

\usepackage{amsmath}	
\usepackage{bm}
\usepackage{graphicx}
\usepackage{amssymb}
\usepackage{amsfonts}
\usepackage{latexsym}
\usepackage{enumerate}
\usepackage{graphicx}
\usepackage{alltt}
\usepackage{array}
\renewcommand{\a}{\alpha}

\usepackage{ulem}
\usepackage{color}

\begin{document}


\title{Reentrant transition in coupled noisy oscillators}


\author{Yasuaki Kobayashi}
\email[]{yasuaki.kobayashi@es.hokudai.ac.jp}
\affiliation{Research Institute for Electronic Science, Hokkaido University, Sapporo 060-0812, Japan}
\author{Hiroshi Kori}
\affiliation{Department of Information Sciences, Ochanomizu University,
Tokyo 112-8610, Japan}


\date{\today}

\begin{abstract}
 We report on a novel type of instability observed in a noisy oscillator
 unidirectionally coupled to a pacemaker. 
 Using a phase oscillator model, we find that, as the coupling strength is increased,
 the noisy oscillator lags behind the pacemaker more frequently and the phase slip rate increases, which may not be observed in averaged phase models such as the
 Kuramoto model. 
 Investigation of the corresponding Fokker-Planck equation
 enables us to obtain the reentrant
 transition line between the synchronized state and the phase slip state. 
 We verify our theory using the Brusselator model, suggesting that
 this reentrant transition can be found in a wide range of limit cycle oscillators.
\end{abstract}


\maketitle

\section{Introduction}\label{introduction}
Synchronization of self-sustained oscillators \cite{winfree01,
kuramoto84, pikovsky01} is crucial in many systems, including cardiac
cells \cite{Tweel, glass01}, circadian clock cells \cite{reppert02,
yamaguchi13}, and power grids \cite{motter, doerfler, timme}. These
systems are inevitably subject to various kinds of perturbations such as
inherent noise, inhomogeneity and environmental changes, and therefore
coupling of such oscillators needs to be strong enough to overcome the
effect of these disturbances and ensure synchronization.

It is known that strong coupling can be a source of instabilities
including oscillation death \cite{ariaratnam01,ermentrout-death} and chaotic dynamics \cite{Parlitz,
kentaro}. However, many of previous studies on coupled oscillators in the presence of noise
focus on the competition of coupling and noise
\cite{pikovsky01, refereeB1,refereeB2, refereeB3}, where coupling is
expected to suppress noise, leading to fast and stable synchronization.

In this paper, we present a new synchronization-breaking scenario which
can occur in the strong coupling regime. 
Naturally, synchronization does not occur for too weak coupling because of the effect of noise. We
find that, in addition to this trivial desynchronization, synchronization is disrupted
also for too strong coupling. Such a
reentrant transition occurs even with very weak noise.
We elucidate the condition and mechanism of this reentrant
synchronization through the analysis of a phase oscillator model.
Furthermore, we verify that the same reentrant transition occurs in limit-cycle oscillators by using
the Brusselator model and confirm the validity of our theory. 
Our study demonstrates that the reentrant transition appears quite generally in coupled noisy oscillators.

\section{Coupled phase oscillator model under noise}
We consider the following phase oscillator that is subject to noise and is influenced by a noise-free pacemaker:
\begin{equation}
 \dot{\phi}=\omega + K Z(\phi)\{h(\Omega t)-h(\phi)\}+\xi(t)\label{eq:phase}, 
\end{equation}
where $\phi$ is the phase, $\omega$ and $\Omega$ are the frequency of the oscillator and the pacemaker, respectively, and
$\xi(t)$ is a Gaussian white noise satisfying
$\langle\xi(t)\xi(t')\rangle=D\delta(t-t')$. 
Interaction is determined by $2\pi$-periodic functions $Z(\phi)$ and $h(\phi)$. 
A large class of limit-cycle
oscillator models can be reduced to this model when the stability of a limit-cycle oscillator is
high enough compared to noise and coupling strengths \cite{kuramoto84}.
Here we adopt the following simple functions: 
\begin{align}
 Z(\phi)=\sin(\phi-\alpha), \quad h(\phi)=-\cos \phi, \label{eq:z_and_h}
\end{align} with a parameter $\alpha$. 
Below we mostly consider the case $\Omega=\omega$. 

\subsection{Averaged model}
Let us first examine the averaged dynamics. When the coupling strength $K(>0)$ and the
noise strength $D$ are sufficiently small compared to $\omega$,
Eq.\eqref{eq:phase} is well approximated by an averaged phase model. When $\Omega = \omega$, 
the phase difference
$\psi\equiv \phi-\omega t$ obeys
\begin{equation}
\dot{\psi}= K\Gamma(\psi)+\xi(t), \label{eq:av_model}
\end{equation}
where the interaction function $\Gamma$ is obtained as \cite{kuramoto84}
\begin{equation}
\Gamma(\psi)\equiv\frac{1}{2\pi}\int_0^{2\pi}d\theta Z(\psi+\theta)\left\{h(\theta)-h(\psi+\theta)\right\}. 
\end{equation}
The
present choice
of $Z$ and $h$ yields a Sakaguchi-Kuramoto type interaction function \cite{sakaguchi86}
\begin{equation}
 \Gamma(\psi)=-\frac{1}{2}\left\{\sin(\psi-\alpha)+\sin \alpha\right\}. 
\end{equation}
In the absence of noise, the state $\psi=0$ is stable when $\Gamma'(0)=-\frac{1}{2}\cos\alpha$ is
negative, i.e., $-\pi/2<\alpha<\pi/2$, which we always consider in the present paper. 

It is convenient to rewrite Eq.\eqref{eq:av_model} as
\begin{equation}
 \dot{\psi}= -K\frac{\partial F(\psi)}{\partial\psi}+\xi(t), \label{eq:av_model_potential}
\end{equation}
where the potential $F$ is given by
\begin{eqnarray}
 F(\psi) &\equiv &  -\int_0^{\psi}d\psi'\Gamma(\psi') \nonumber\\
 &= & -\frac{1}{2}\cos(\psi-\alpha)+\frac{1}{2}(\sin\alpha)\psi.
\end{eqnarray}
If $K \gtrsim D$, the phase difference $\psi$ tends to stay around the potential minima, and
occasionally jumps to the two adjacent minima, driven by noise. If $\alpha=0$, the right and the
left potential barriers are the same height, and no net drift appears. On the other hand, if $\alpha
\neq 0$, the imbalance of the two barriers causes nonzero drift in the positive direction for
$\alpha<0$, or in the negative direction for $\alpha>0$. The rate of phase slip is well approximated by
Kramers' formula \cite{pikovsky01}: Given a barrier height $\Delta F$, the rate of overcoming the
barrier is proportional to $\exp(-\frac{K\Delta F}{D})$. Since the difference between the right and
the left barriers is $\pi\sin\alpha$, the net jump rate is proportional to 
\begin{equation}
\exp\left(-\frac{\Delta F_{\mathrm{L}} K}{D}\right)\left\{
\exp\left(-\frac{K\pi\sin\alpha}{D}\right)-1 \right\},
\end{equation}
where $\Delta F_{\mathrm{L}}$ is the height of the left barrier. That is, in the averaged dynamics,
the rate of phase slip decreases exponentially as $K$ increases, with its direction determined by the
sign of $\alpha$, and no reentrant transition occurs. 

\subsection{Non-averaged model}
\begin{figure*}[ht]
\includegraphics[width=.7\linewidth]{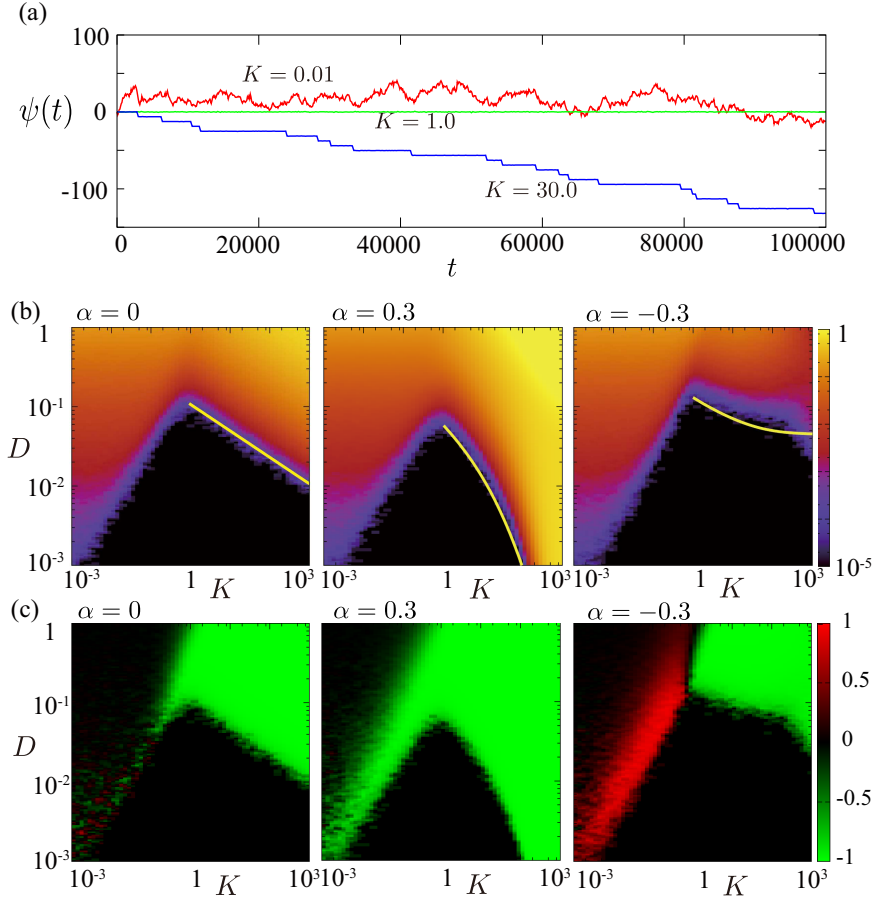}
\caption{(Color online) (a) Time evolution of the phase
 difference $\psi$ for three different values of $K$, with $\alpha=0$ and $D=0.05$.  (b) Phase slip
 rate against the coupling strength $K$ and the noise intensity $D$ for $\alpha=0$, $0.3$, and
 $-0.3$. Theoretical lines are given by Eqs.\eqref{eq:scaling2-1} and \eqref{eq:scaling2-2} for $K\ge 1$. (c)
 Unidirectionality of slip against $K$ and $D$. Red and green indicate positive and negative
 directions of $\psi$, respectively. 
}\label{fig:phase_diag}
\end{figure*}

When $K$ or $D$ is not sufficiently smaller than $\omega$, the averaging is no longer valid. In this case, the situation changes drastically.
To see this, we numerically solve Eq.~(\ref{eq:phase}). The frequency $\omega$ can be set to unity
without loss of generality. 
Figure \ref{fig:phase_diag}(a) shows three
types of behavior for different values of $K$ with $D=0.05$ and
$\alpha=0$: For $K=0.01$, the system is dominated by
noise (incoherent region).
Synchronization is observed when $K$ is increased to
$K=1.0$. However, as we further increase $K$ up to $K=30$,
synchronization is disturbed by a jump of the phase
difference by $-2\pi$ (i.e., a phase slip). 
Measuring the rate of phase slip events against $K$ and $D$ for different values of $\alpha$, we observe the
reentrant transition from the synchronized state to the phase slip
state [Fig.\ref{fig:phase_diag}(b)]. 
The critical value of $K$ for this transition decreases as $D$ increases.

Unidirectionality of the slips is seen in Fig. \ref{fig:phase_diag}(c). 
Here, for each $K$ and $D$,
we plot $(N_{+}-N_{-})/(N_{+}+N_{-}+\epsilon)$, where $N_{+}$ and $N_{-}$ are the number of slips in the
positive and the negative directions, respectively, and $\epsilon=1$ is inserted to
circumvent zero division.
In the incoherent region, 
the slip direction and frequency is determined by the sign of $\alpha$, reflecting the drifting
force $-\frac{K}{2}\sin\alpha$ appearing in the averaged dynamics. On the other hand, in the reentrant
region, the slip direction is negative for all $\alpha$ values, implying that the phase slip in
the non-averaged dynamics is qualitatively different from that in the averaged dynamics.

The local stability analysis cannot explain the reentrant transition. Linearizing
Eq.\eqref{eq:phase} around the synchronized state $\psi = 0$ with
$\omega=\Omega$ yields
\begin{equation}
 \dot{\psi}=-K\lambda(\omega t)\psi + \xi(t), 
\end{equation}
where $\lambda(\omega t) \equiv Z(\omega t)h'(\omega t)$. 
The linear stability of the state $\psi=0$ is determined by the time average of $\lambda(\omega t)$,
i.e., $(\omega/2\pi)\int_0^{2\pi/\omega} \lambda(\omega t) dt = - (1/2) \cos \alpha$, which is the
same as the stablity in the averaged model. The state $\psi=0$ is thus linearly stable for $-\pi/2 <
\alpha < \pi/2$. 
However, it should be noted that $\lambda(\omega t)$ can be negative for some range of time even
when its time average is positive. Hence it seems likely that the system is more easily destabilized
in the non-averaging model than the averaging one. Nevertheless, the reentrant transition is observed
even when there is no time interval for the coefficient to be negative. In fact, considering the case
$\alpha=0$, where $\lambda(\omega t)=\sin^2\omega t \ge 0$, the synchronized state is never
destabilized at any time, which suggests that nonlinearity is responsible for the reentrant transition. 

\begin{figure}[tb]
\includegraphics[width=.8\linewidth]{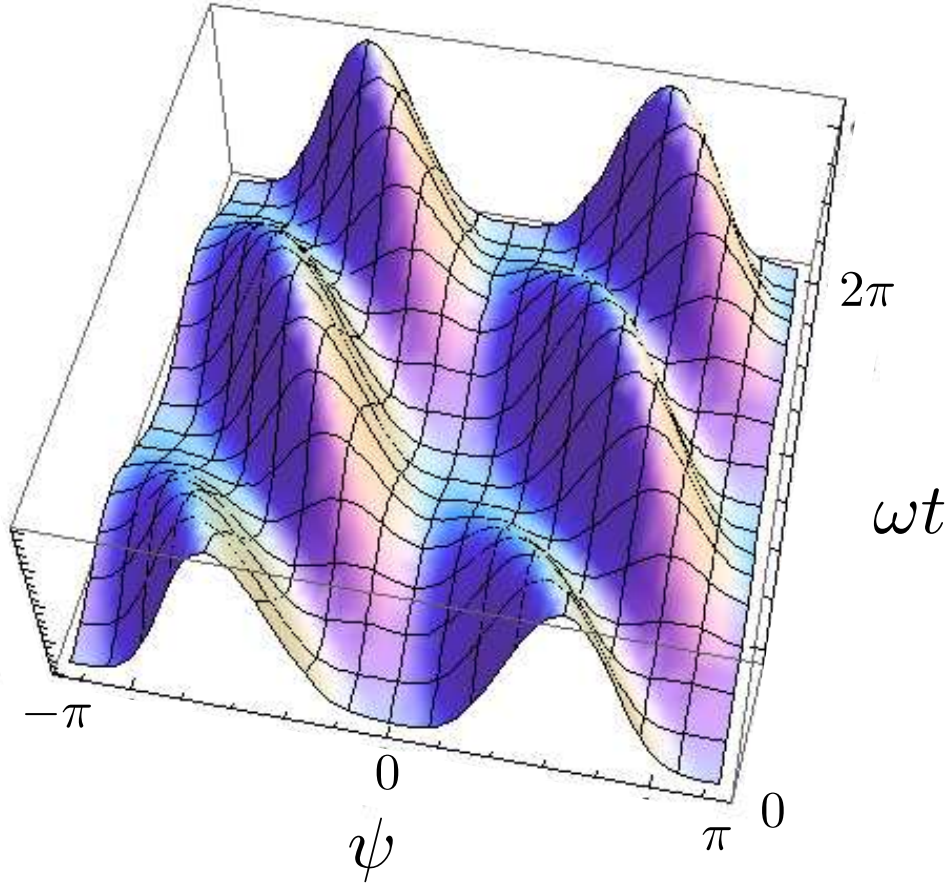}
\caption{(Color online) Space-time plot of the time dependent potential $F(\psi,t)$ (Eq.(\eqref{eq:pot_explicit})) for $\alpha=0$. }\label{fig:potential}
\end{figure}

To understand the global structure of the system, we again utilize a potential description.
Equation\eqref{eq:phase} can be rewritten as
\begin{equation}
 \dot{\psi}=-K\frac{\partial F(\psi,t)}{\partial \psi} + \xi(t),
\end{equation}
where 
\begin{align}
 F(\psi,t) \equiv &-\int_0^{\psi}\!\!\!d\psi' Z(\psi'+\omega t)\left\{h(\omega
t)-h(\psi'+\omega t)\right\} \\
=& -\frac{1}{2}\cos(\psi -\alpha) - \frac{1}{2}\cos(\psi+2\omega t-\alpha) \nonumber\\
&+\frac{1}{4}\cos (2\psi+2\omega t- \alpha) + \frac{1}{2}(\sin\alpha)\psi \nonumber \\
&+\frac{1}{2}\cos\alpha+\frac{1}{4}\cos(2\omega t -\alpha). \label{eq:pot_explicit}
\end{align}
Note that the potential $F(\psi,t)$ is now time dependent. 
In general, $F(\psi,t)$ is a
2$\pi$-periodic function in $\omega t$. Choosing $Z$ and $h$ as in Eq.\eqref{eq:z_and_h}, $F$ is $\pi$-periodic in $\omega t$. 
Figure \ref{fig:potential} shows the space-time plot of $F(\psi,t)$ for $\alpha=0$. It is clearly
seen that, in addition to the minimum $\psi=0$, which exists in the case of the averaged dynamics as
well,
there is another minimum traveling in the negative direction of $\psi$. One can easily confirm that
$F(\psi,t)$ has the traveling minimum $\psi=-2\omega t$ and the two maxima $-\omega t + \alpha$ and
$-\omega t+\alpha + \pi$. It is expected that, in the
non-averaged dynamics, phase slip occurs along this traveling minimum. 

The potential is not tilted if $\alpha=0$, and tilted if $\alpha\neq 0$. Since the former is easier to
analyze, below we first present our analysis for $\alpha=0$, and then extend the analysis to the case
$\alpha\neq 0$. 

\subsubsection{Non-tilted potential ($\alpha=0$)}
\begin{figure}[tb]
\includegraphics[width=\linewidth]{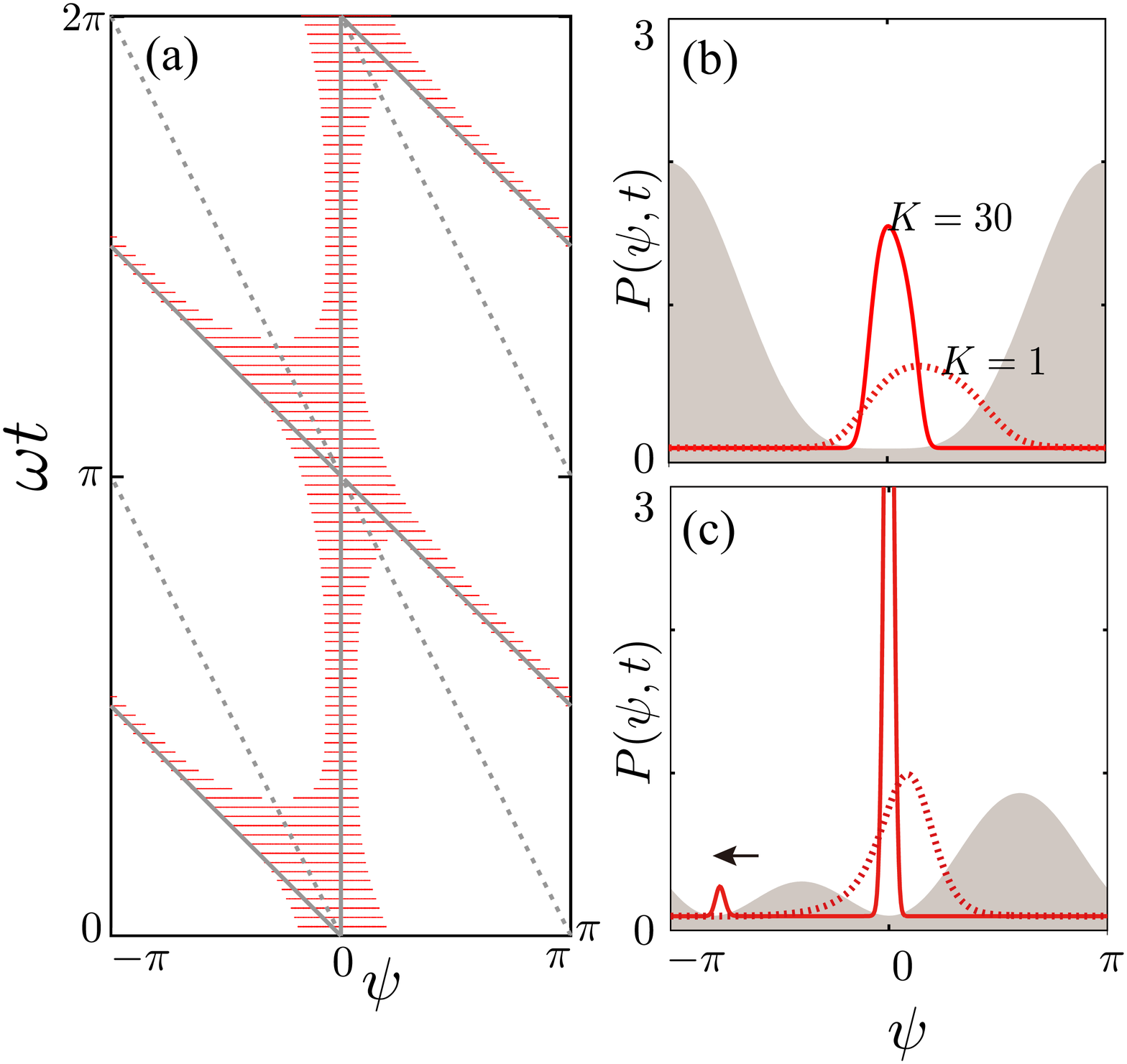}
\caption{(Color online) (a) Space-time plot of the probability distribution function $P(\psi,t)$ obtained by
 solving Eq.\eqref{eq:fokker-planck} with $K=30$, $D=0.1$, and $\alpha=0$. Only the region $P(\psi,t)>0.01$ are shown (red).  The minima (solid lines) and the maxima (dotted
 lines) of the potential $F(\psi,t)$ are also shown. (b, c) Snapshots of $P(\psi,t)$ for $K=1$ (dotted) and $K=30$ (solid), with $D=0.1$ and
 $\alpha=0$. Shaded regions represent $F(\psi, t)$. (b) $t=0$, (c) $t=1.25$. Probability current corresponding to the phase slip is indicated by an
 arrow in (c). 
}\label{fig:snapshots}
\end{figure}

The mechanism and condition of the phase slip can be understood 
through the analysis of the following Fokker-Planck equation:
\begin{equation}
\frac{\partial P(\psi,t)}{\partial t}=K\frac{\partial}{\partial \psi}
\left(\frac{\partial F}{\partial \psi}P\right)+\frac{D}{2}\frac{\partial^2 P}{\partial \psi^2}.
\label{eq:fokker-planck}
\end{equation}
where $P(\psi,t)$ is the probability distribution function for the phase difference $\psi$. 
By numerically solving Eq.\eqref{eq:fokker-planck} with Eq.\eqref{eq:pot_explicit} for $\alpha=0$, we obtain time
evolution of $P(\psi,t)$. Figure \ref{fig:snapshots}(a) shows $P(\psi,t)$
together with the trajectories of the extrema of $F(\psi,t)$. Note that, for $\alpha=0$, three of
them cross each other at the same time. When the coupling is
sufficiently large, $P$ splits into two components at some moment,
one localized at $\psi=0$ corresponding to the synchronized state,
and the other traveling along
$\psi=-2\omega t$, which corresponds to the phase slip state. 
This traveling component appears only for sufficiently strong coupling,
although the distribution around $\psi=0$ is sharper for stronger coupling [Fig.\ref{fig:snapshots}(b),(c)].

The scenario of how the traveling component emerges is as follows. Let us focus on a short time
interval around $t=0$. When $t<0$, $P$ is localized at $\psi=0$.
At $t=0$, the three extrema of $F$ cross each other [Fig.~\ref{fig:snapshots}(a)]. Around this time, the
curvature of the potential at $\psi=0$ almost vanishes, and hence the
diffusion dominates the dynamics. 
Then, for $t>0$, a potential maximum located at $\psi=-\omega t$ gradually grows,
and, at some time $t_c$ the drift force caused by the potential becomes comparable to the effect of diffusion. 
At this moment, the part of $P(\psi,t)$ located beyond the maximum (i.e., $\psi<-\omega
t$) is separated from the component around $\psi=0$ and thus conveyed with the
potential minimum located at $\psi=-2\omega t$. This process repeats
itself with period $\pi$. 

We can roughly estimate the above $t_c$ and the transition line through a dimensional
analysis. First, $t_c$ is determined by balancing the drift and the diffusion terms
in Eq.~\eqref{eq:fokker-planck} evaluated at the potential maximum $\psi=-\omega t$:
\begin{equation}
K F(-\omega t_c, t_c)\sim D. \label{eq:cond1}
\end{equation}
Because diffusion is dominant for $0<t<t_c$, the width of $P$ 
grows roughly as $\sqrt{Dt}$ within this duration. If this width is comparable to
distance to the potential maximum at $t=t_c$, i.e., 
\begin{equation}
\sqrt{Dt_c}\sim \omega t_c, \label{eq:cond2}
\end{equation}
a substantial part of $P$ will be
conveyed.
From Eqs. \eqref{eq:cond1} and \eqref{eq:cond2}, and noting that the
potential height for small $t$ is $F(-\omega t,t)\sim (\omega t)^4$,  we obtain the following scaling relation:
\begin{equation}
D\sim \omega^{\frac{4}{3}}K^{-\frac{1}{3}}. \label{eq:scaling1}
\end{equation}
By substituting Eq.~\eqref{eq:scaling1} into Eq.~\eqref{eq:cond2}, we obtain the relationship
between $K$ and $t_c$:
\begin{equation}
t_c\sim \omega^{-\frac{2}{3}}K^{-\frac{1}{3}}. \label{eq:scalingtc}
\end{equation}
The reason why stronger coupling induces more phase slips
is now clear: Since larger coupling strength $K$ implies smaller $t_c$, the condition
$\sqrt{Dt_c}>\omega t_c$ is easier to satisfy. 

\subsubsection{Tilted potential ($\alpha\neq 0$)}
\begin{figure}[tb]
\includegraphics[width=\linewidth]{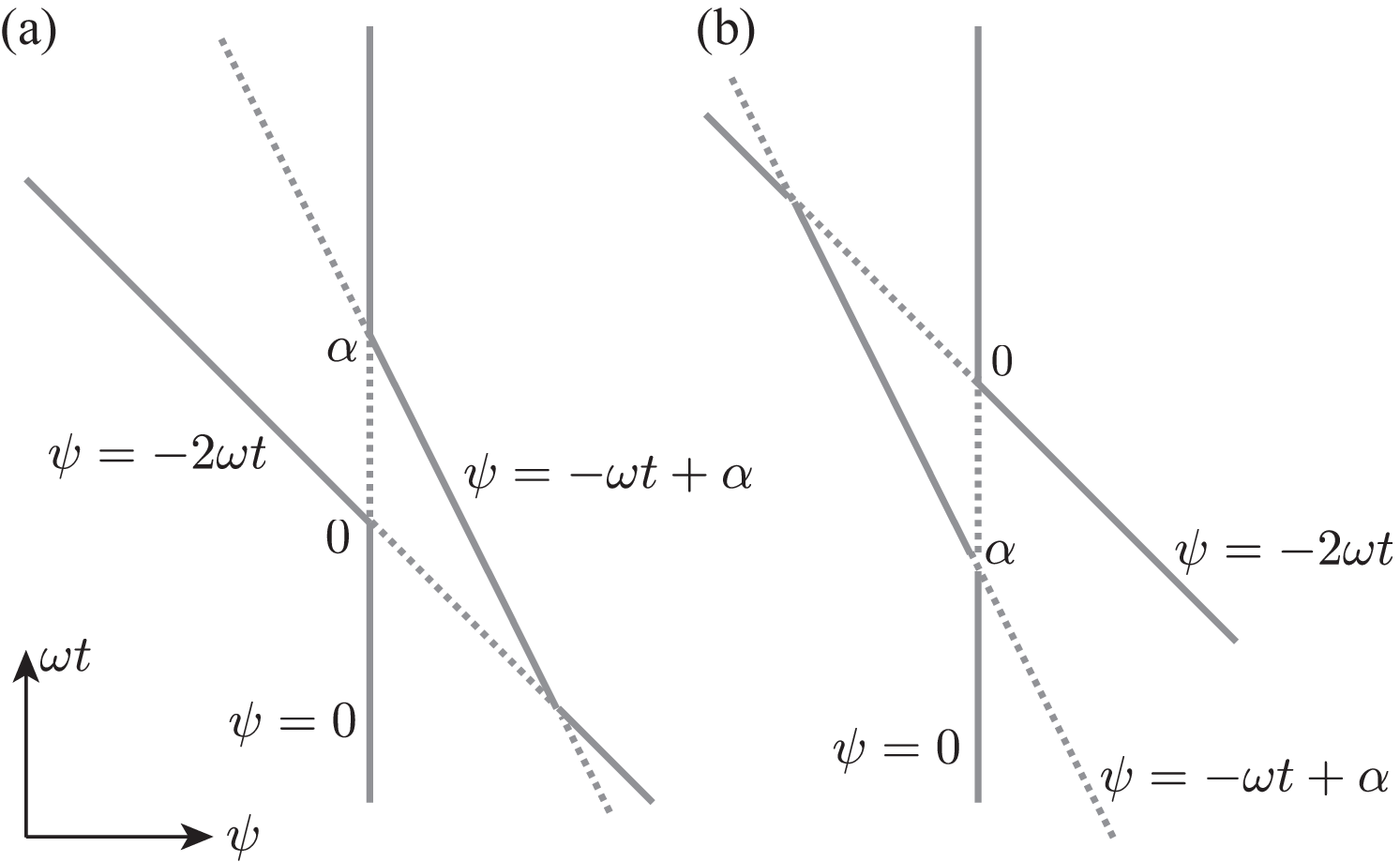}
\caption{Schematic of potential extrema crossing for nonzero $\alpha$: (a)$\alpha>0$, (b)$\alpha<0$. Solid (dotted) lines correspond to minima (maxima) of the potential $F(\psi,t)$. }\label{fig:schematic}
\end{figure}

In the case of small but nonzero $|\alpha|$, where the three extrema cross
each other at different timings, a similar argument can still be made
to obtain the transition lines: If $\alpha>0$, $\psi=0$ becomes unstable
when $\psi=-2 \omega t$
crosses 0 at $t=0$.  On the
other hand, if $\alpha<0$, $\psi=0$ loses its stability at
$t=-\omega^{-1}|\alpha|$, when $\psi=-\omega t+\alpha$ crosses $\psi=0$.  In both
cases, the potential barrier is located at $\psi=-\omega t +
\alpha$. 
These situations are schematically shown in Fig.~\ref{fig:schematic}.
Hence, instead of Eq.\eqref{eq:cond2}, we have different
balance equations $\lambda\sqrt{Dt_c}=\omega t_c-\alpha$ ($\alpha\ge
0$) and $\lambda\sqrt{D(t_c-\omega^{-1}\alpha)}=\omega t_c-\alpha$
($\alpha<0$), where a parameter $\lambda$ has been introduced. 
The transition line is thus obtained by eliminating $t_c$ from the following
equations:
\begin{gather}
K F(-t_c+\alpha, t_c)=D , \label{eq:scaling2-1}\\
\omega t_c=\left\{
\begin{array}{l}
\frac{\omega^{-1}D\lambda^2}{2}+\a + \sqrt{\left(\frac{\omega^{-1}D\lambda^2}{2}+\a\right)^2-\a^2} \quad (\a \ge 0), \\
\lambda^2 \omega^{-1}D +\a \quad (\a < 0)
\end{array}
\right.\label{eq:scaling2-2}
\end{gather}
Substituting $\alpha=0$ obviously reproduces Eqs.\eqref{eq:cond1} and \eqref{eq:cond2}. Theoretical
lines given by Eqs.\eqref{eq:scaling2-1} and \eqref{eq:scaling2-2} with $\lambda=3.0$ agree well with numerical data for $\alpha=0$ and $\alpha=0.3$ in Fig.~\ref{fig:phase_diag}(b), although for $\alpha=-0.3$ it
slightly deviates from the numerical data particularly for large $K$. 
This result suggests that the sign of $\alpha$ is critial for stable
synchronization: phase slip hardly occurs if $\alpha<0$ even for large $K$.

\subsubsection{Frequency mismatch}
\begin{figure}[tb]
\includegraphics[width=\linewidth]{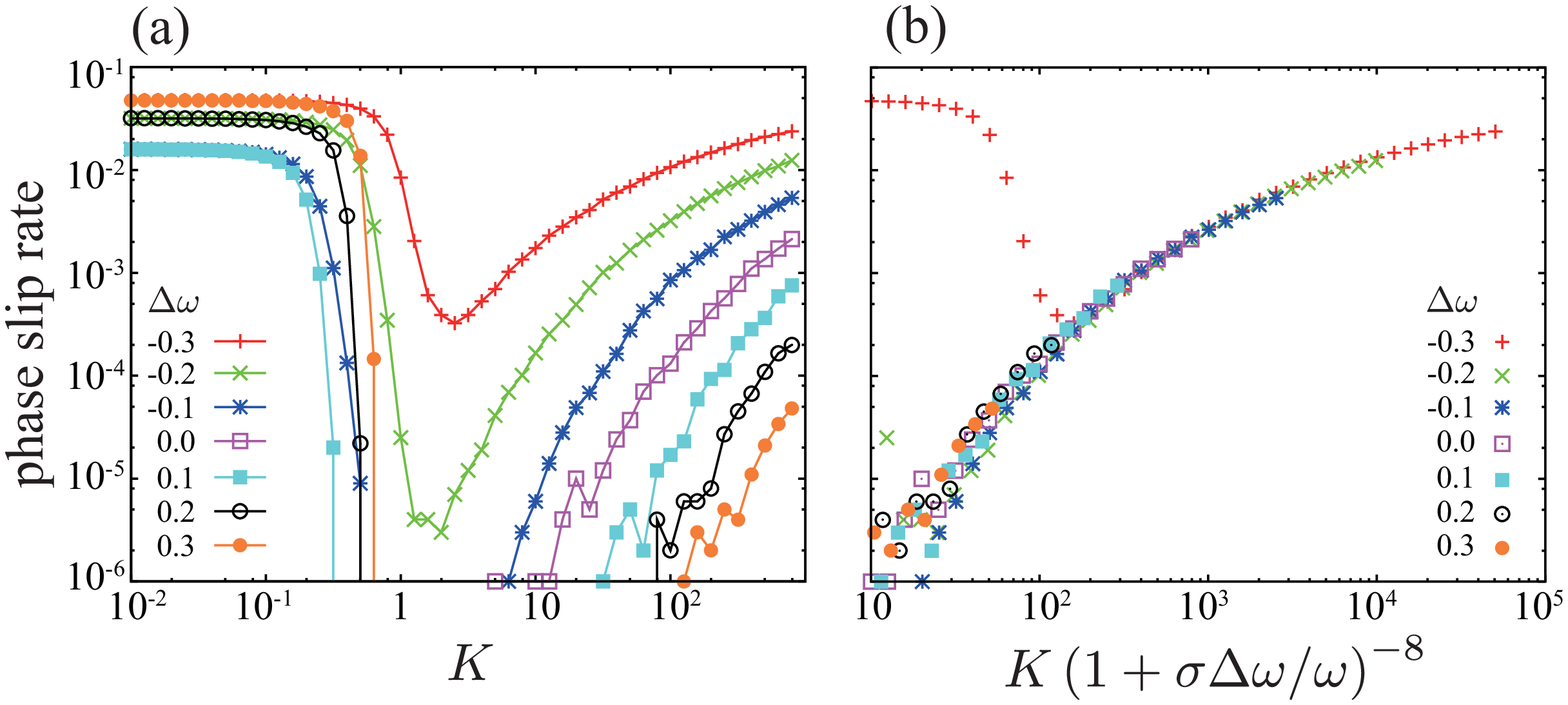}
 \caption{(Color online) (a) Phase slip rates against $K$ for different $\omega=\Delta\omega+\Omega$, with
 $D=0.3$, $\alpha=0$ and $\Omega=1$. (b) Same data plot for larger $K$, scaled by
 $(1+\sigma\Delta\omega/\omega)^{8}$, with the fitting parameter
 $\sigma$=1.35. }\label{fig:freq_diff}
\end{figure}

We can further extend our scaling argument for $\alpha=0$ 
to the case $\Omega \neq \omega$.
Since the frequency difference $\Delta \omega=\omega-\Omega$ serves as
a drifting force, Eq.\eqref{eq:cond2} is now modified as
\begin{equation}
 \lambda\sqrt{Dt_c}-\sigma\Delta\omega t_c = \omega t_c, 
  \label{eq:with_deltaomega}
\end{equation}
where $\sigma>0$ is another parameter. Combining Eq.~\eqref{eq:with_deltaomega}
with Eq.\eqref{eq:cond1}, we obtain 
\begin{equation}
 D\sim
  \omega^{\frac{4}{3}}\left(1+\sigma\frac{\Delta\omega}{\omega}\right)^{\frac{8}{3}}K^{-\frac{1}{3}}. 
  \label{eq:scaling3}
\end{equation}
Figure \ref{fig:freq_diff}(a) shows the phase slip rates for different values of
$\Delta\omega$. Rescaling according to Eq.~\eqref{eq:scaling3} results in the collapse of data points
[Fig.\ref{fig:freq_diff}(b)], which strongly supports the validity of our theory.

\subsection{Generality of reentrant transition}
\label{sec:general}
We have studied the non-averaged phase model \eqref{eq:phase} using a specific form of $Z$ and $h$
as in Eq.\eqref{eq:z_and_h}. Now let us consider reentrant transition in more general settings.  For
simplicity, we again focus on the case $\Omega=\omega$. As we have observed above, the extrema of
the potential $F(\psi,t)$, which are given by zeroes of $Z(\psi+\omega t)$ and $H(\psi,t)\equiv
h(\omega t)-h(\psi+\omega t)$ for $0\le \psi \le 2\pi$, play important roles on phase slip.
Obviously $H$ has a zero at $\psi=0$, which corresponds to the synchronized state. Now we assume
that the functions $Z$ and $h$ are unimodal and that $Z$ has two zeroes, which are reasonable in
many practical cases.  
Then it follows that $H$ has another zero $\psi=\beta(t)$, which we call the
moving $H$-branch, corresponding to phase slip. In fact, from the above assumption it is easy to
show that $\beta(t)$ is a monotonically decreasing function of $t$ that satisfies
$\frac{d\beta(t)}{dt}<-\omega$, crossing $\psi=2n \pi$ ($n=0,1,2,\cdots$) twice within one
oscillation period, reflecting the fact that phase slips occur in the negative direction. In
addition, two zeros of the response function $Z(\psi+\omega t)$, denoted by $\alpha_1$ and
$\alpha_2$, give the other potential extrema moving linearly along $\psi=-\omega t+\alpha_{1,2}$,
which we call $Z$-branches. Note that, in the case of Eq.\eqref{eq:z_and_h}, the $Z$-branches are $\psi=-\omega t+\alpha$ and $-\omega t+\alpha+\pi$, and
the moving $H$ branch is $\psi=-2\omega t$.

The synchronized state $\psi=0$ is destabilized by the crossing of the $Z$-branches and the
moving $H$-branches, and phase slip may occur, depending on in which order these branches cross
$\psi=0$: if the branch crossing is like Fig.\ref{fig:schematic}(a), that is, if $\psi=0$ is
destabilized by the moving $H$ branch first, and then re-stabilized by a $Z$-branch, phase slip is
more likely to occur; if the crossing is like Fig.\ref{fig:schematic}(b), phase slip is less likely. 

\section{Reentrant transition in coupled limit cycle oscillators}
Let us confirm that the reentrant transition occurs in limit cycle oscillators. We demonstrate it by
using the Brusselator model:
\begin{eqnarray}
 \dot u_i &=& A -(B+1) u_i + u_i^2 v_i+K^{(u)}_{i}(u_j-u_i)+\xi_i(t), \label{eq:brussel_u}\\
 \dot v_i &=& Bu_i-u_i^2 v_i+K^{(v)}_{i}(v_j-v_i)+\eta_i(t),\label{eq:brussel_v}
\end{eqnarray}
where $(i,j)=(1,2)$ or $(2,1)$, $u_{i,j}$ and $v_{i,j}$ are the state variables, and $\xi_i(t)$ and $\eta_i(t)$ are
the Gaussian white noise satisfying $\langle \xi_i(t)\xi_i(t')\rangle=\langle
\eta_i(t)\eta_i(t')\rangle=D_i\delta(t-t')$. 
Here we consider the case where oscillator 2 acts as a noise-free pacemaker: $D_2=0$, 
$K^{(u)}_{2}=K^{(v)}_{2}=0$. On the other hand, oscillator 1 is under noise, $D_1=D$, and is
influenced by oscillator 2 via $u$-coupling ($K^{(u)}_{1}=K$, $K^{(v)}_1=0$) or $v$-coupling
($K^{(u)}_{1}=0$, $K^{(v)}_1=K$). 

\begin{figure}[tb]
\includegraphics[width=\linewidth]{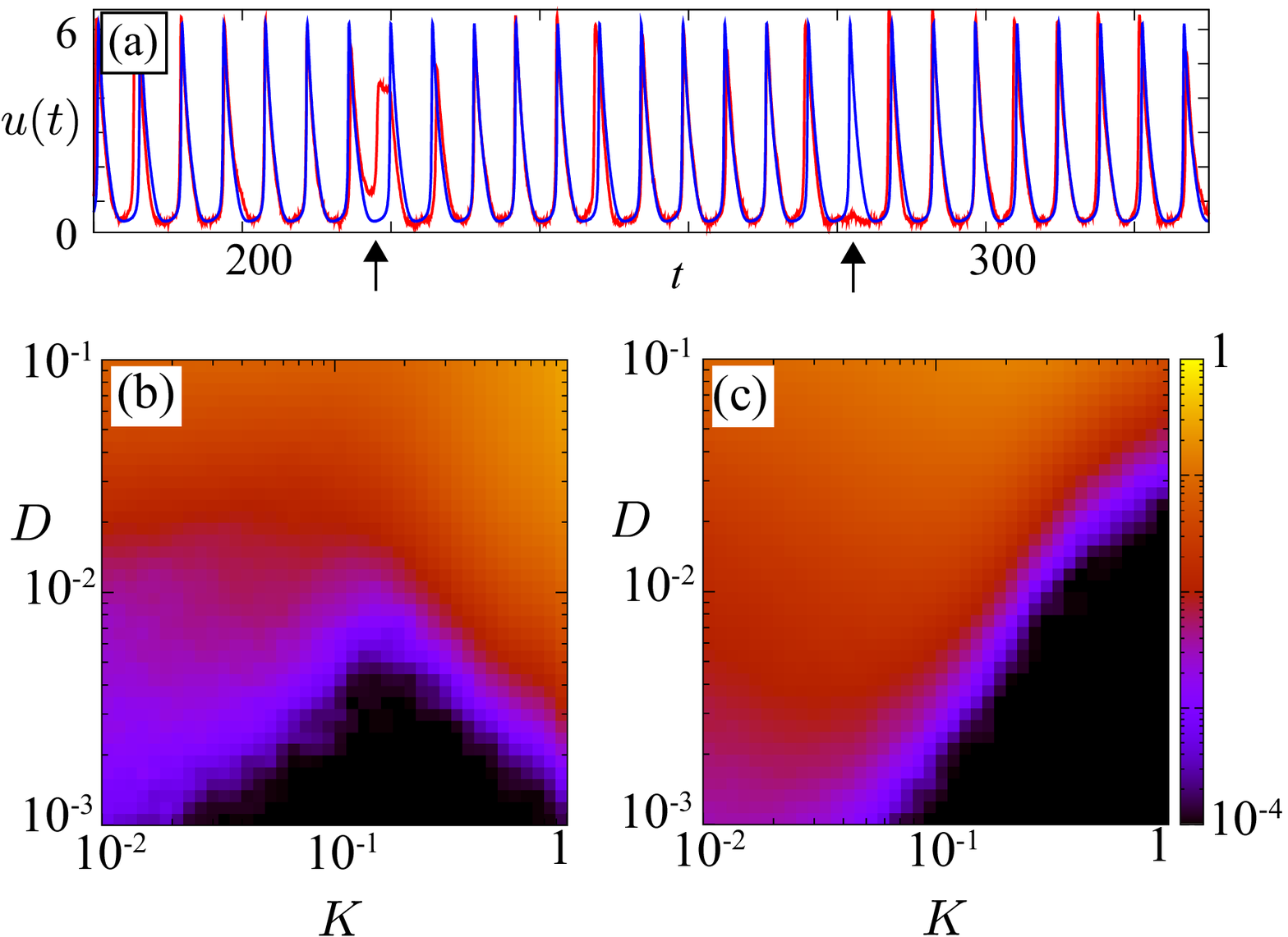}
 \caption{(Color online) Two unidirectionally coupled Brusselators: 
Parameters are $A=1.6$ and $B=5.0$. (a) Time series of $u_1(t)$ (red) and $u_2(t)$
 (blue) in the case of $v$-coupling, with $K=0.5$ and $D=0.01$. Oscillation failures are observed at $t\sim 220$ and $t\sim$ 280,
 indicated by arrows. 
(b, c) Frequency of phase slip against $D$ and $K$: (b)$v$-coupling, (c)$u$-coupling.
}\label{fig:brussel}
\end{figure}

In the case of $v$-coupling, too strong coupling results in occasional oscillation failures, as
shown in Fig.\ref{fig:brussel}(a). If we measure the phase of the state $(u,v)$ in a standard
way \cite{kuramoto84}, oscillation failure can be expressed as phase slip. The frequency of
phase slips increases as $K$ increases for fixed $D$, indicating a reentrant transition
from synchronization to phase slip [Fig.\ref{fig:brussel}(b)].  On the other hand,
$u$-coupling does not induce the reentrant transition [Fig.\ref{fig:brussel}(c)]. 

\begin{figure}[t]
\includegraphics[width=\linewidth]{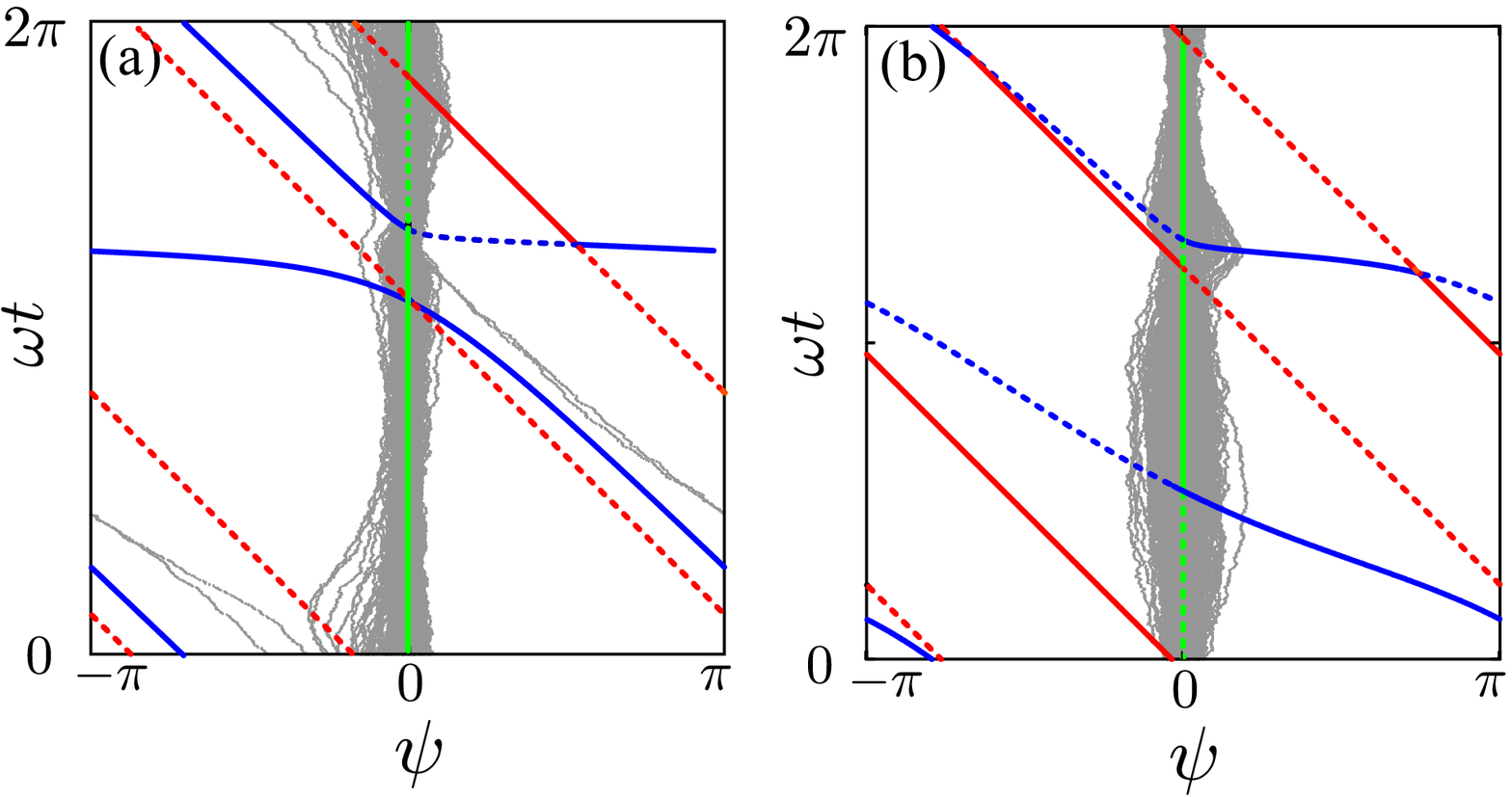}
 \caption{(Color online) Trajectories (gray) of the phase difference $\psi(t)$ obtained by solving Eq.\eqref{eq:phase}, with $Z$ and $h$ calculated from the numerical phase
reduction of Eqs.\eqref{eq:brussel_u} and
\eqref{eq:brussel_v}, with
 $D=0.005$, $K=1.0$. The extrema of the numerically constructed potential
 $F(\psi,t)$ are shown together: $\psi=0$ (green), $Z$-branches (red), and a moving $H$-branch (blue). Solid and dotted lines correspond to minima and maxima of the potential, respectively. The trajectory is
 plotted modulo $2\pi$ in $\omega t$. (a)$v$-coupling, (b)$u$-coupling. }\label{fig:zerolines}
\end{figure}

The difference between the two ways of coupling can be understood by applying the analysis in
Sec.~\ref{sec:general} to the phase model of the Brusselator.  
We numerically solve Eq.\eqref{eq:phase} with $Z$ and $h$ calculated from the numerical phase
reduction of Eqs.\eqref{eq:brussel_u} and
\eqref{eq:brussel_v}.  
The trajectory of the phase difference $\psi(t)$ and
the potential extrema are shown in Fig.\ref{fig:zerolines}. 
Here we can clearly see the two types of branch crossing discussed in Sec.~\ref{sec:general}: 
In the case of $v$-coupling
[Fig.\ref{fig:zerolines}(a)], it is
observed that $\psi(t)$ occasionally leaves the 
state $\psi=0$ and
travels along the moving $H$-branch.  This occurs when the state $\psi=0$ becomes
temporally unstable due to the crossing by the moving $H$-branch. Its
stability recovers when the state $\psi=0$ is crossed by one of the
$Z$-branches. This branch crossing is the type $\alpha > 0$ [Fig.\ref{fig:schematic}(a)].  On the
other hand, in the case of $u$-coupling, phase slip is not observed [Fig.\ref{fig:zerolines}(b)],
where branch crossing is of the type $\alpha <0$ [Fig.\ref{fig:schematic}(b)]. 
These results correspond to the presence and the absence of the reentrant transition in
Fig.\ref{fig:brussel}(b) and (c), respectively.

It is remarkable
that the reentrant transition in the Brusselator model is well
understood from the analysis of its corresponding phase model for such a
strong coupling case. Our results indicate that non-averaged phase
models are useful for the understanding of strongly coupled
oscillators. 

\section{Concluding remarks}
Our study has uncovered that, in addition to a lower limit in coupling strength,
there is generally an upper limit over which synchronization is disrupted by
phase slips when an oscillator is subject to noise. 
Therefore, when strong coupling is required, the way of coupling should carefully be constructed.

In this study we have only considered unidirectional coupling. In the case of mutual coupling, we
have confirmed in a preliminary numerical study that mutual coupling also exhibits reentrant
transition, although oscillation death is more commonly observed. 

A similar eentrant transition is known to occur in a certain class of chaotic oscillators
\cite{heagy-pecora, huang-pecora}. Our study would further motivate such studies.

\begin{acknowledgments}
We thank Dr. Fumito Mori for valuable discussions. 
\end{acknowledgments}


\end{document}